\documentclass[letterpaper,12pt]{article}
\usepackage{amssymb,latexsym,amsmath,amsthm}

\makeatletter \@addtoreset{equation}{section} \makeatother

\input epsf

\usepackage{pdfsync}
\usepackage{fullpage}

\begin{document}
\begin{titlepage}
 \thispagestyle{empty}
\begin{flushright}
     \hfill{CERN-PH-TH/2010-126}\\
     \hfill{SU-ITP-10/20}
 \end{flushright}

 \vspace{50pt}

 \begin{center}
     { \Large{\bf      {Small N=2 Extremal Black Holes in Special Geometry
    }}}

     \vspace{75pt}

     {\large { Anna Ceresole$^{a}$, Sergio Ferrara$^{b,c}$ and Alessio Marrani$^{d}$ }}

     \vspace{50pt}

   { \small  {\it ${}^a$ INFN, Sezione di Torino,
       Via Pietro Giuria 1, 10125 Torino, Italy} 

     \vspace{10pt}

  {\it ${}^b$ Physics Department, Theory Unit, CERN,\\
     CH -1211, Geneva 23, Switzerland}}

     \vspace{10pt}

  {\small  {\it ${}^c$ INFN - Laboratori Nazionali di Frascati,\\ 
     Via Enrico Fermi 40, I-00044 Frascati, Italy}  }

     \vspace{10pt}

  {\small  {\it ${}^d$ Stanford Institute for Theoretical Physics,\\ 
     Stanford University, Stanford, CA 94305-4060,USA}  }

     \vspace{15pt}

     \vspace{100pt}

     {ABSTRACT}
 \end{center}

 \vspace{10pt}
\noindent
We provide an intrinsic classification of the large and small orbits for N=2, 4D extremal black holes on symmetric spaces which does not depend on the duality frame used for the charges or on the special coordinates. A coordinate independent formula for the fake superpotential W, which (at infinity) represents the black hole ADM mass, is given explicitly in terms of  invariants of the N=2 special geometry.
\vfill

\end{titlepage}

\baselineskip 6 mm

\section{Introduction} \label{Intro}
Black hole solutions of (super-)gravity theories with abelian vector and scalar fields, arising at low energy from superstring or M-theory, are presently at the centre of  a broadening field of research. In virtue of their attractor behaviour \cite{SUSY,FGK}, extremal, static, spherically symmetric black holes have been analyzed  for N-extended supergravities in various space-time dimensions, both in their BPS and the non-BPS branch \cite{larsen,varinonBPS}.

 The main properties of these black holes are encoded in the effective potential of the geodesic action which, in N=2 language, is given by
\begin{equation}
	V_{BH}=|Z|^2 +|D_i Z|^2\,,\label{pot}
\end{equation}
where $Z(\varphi,\bar \varphi ;q,p)$ is the N=2 central charge \cite{SUSY,FGK}. The extrema of this potential with respect to the scalar fields, $\partial_\varphi V_{BH}=0$ yield the attractor solutions, while the attractor values of $V_{BH}$ give  the corresponding entropies. In turn, the effective potential  can be written in terms of a real scalar function $W(\varphi^i,\bar \varphi^i;q,p)$ by 
\begin{equation}
V_{BH} = W^2+4g^{i{\bar\jmath}}\partial_i W \partial_{\bar\jmath} W\, \label{VW}\, .
\end{equation}
$W$  governs the first order flow (BPS) equations for the radial evolution of the complex scalar fields $\varphi^i$ and the warp factor $U$ from asymptotic infinity towards the black hole horizon :
\begin{equation}
U^\prime =-e^UW\, , \qquad \qquad \varphi^{\prime i}=-2e^Ug^{i{\bar\jmath}}\partial_{\bar\jmath} W\, . \label{flows}
\end{equation}
The superpotential $W$ also encodes many basic characteristics of the extremal black hole, such as the horizon entropy $S_{BH}=\pi W^2$,  the $ADM$ mass $M_{ADM}=W$ at infinity, and the scalar charges at infinity, $\Sigma_i=\partial_i W$.
For BPS solutions,  $W$ is given by the modulus of the central charge $|Z|$ according to (\ref{pot})\cite{SUSY}. The BPS flows end at the critical points of the central charge $D_iZ=(\partial_i+\frac12 \partial_i K)Z=0$, which are also critical point of the full potential (\ref{VW}).
However \cite{Ceresole:2007wx}, the same set up  for the non BPS branch\cite{larsen} requires that one identifies  a  {\sl fake} superpotential $W(\varphi,\bar\varphi ; q,p)$ whose  extrema now describe the non supersymmetric attractors. Therefore,  full  solutions are more readily obtained from first order BPS-like flow equations  as in the supersymmetric case.  For these reasons, computing  W has been  recently a topical issue. 

It has been known for some time that the U-duality of the underlying supergravity dictates many important features of the solutions \cite{Cvetic:1996zq}. This is perfectly illustrated by the maximally extended  $N=8$ theory, where there are 70 scalar fields spanning the symmetric space $E_{7(7)}/SU(8)$ and the pointlike electric-magnetic charges form a 56-dimensional vector $Q^a$ of the U-duality group $E_{7(7)}$. The area of the horizon for both 1/8-BPS and non-BPS attractors is proportional to $\sqrt{\pm I_4}$, where $I_4=T_{abcd}Q^aQ^bQ^cQ^d$ is the $E_{7(7)}$ quartic invariant, while a positive or negative sign in $I_4$ tells apart a BPS from  a non-BPS attractor \cite{Kallosh:1996uy}. 
More generally, it was shown in \cite{Ferrara:1997ci} that for a generic  U-duality group G in extended supergravity, different supersymmetry features of the p-brane solutions can be characterized by some G-invariant conditions on the central charges. 
For fixed values of the $I_4$ invariant in 4d and of  analogous cubic invariants $I_3$ existing in 5d, the charge vector Q for supergravity theories based on symmetric spaces  describes orbits whose nature determines the amount of supersymmetry preserved by  the attractor points \cite{FG}. This leads to a classification of the attractors in terms of orbits of the charge vector Q \cite{orbits,orbits2}.  

Orbits subdivide into regular or ``Large'', and singular or `` Small''. Small orbits arise for vanishing horizon area, when $I_4=0$, they  have zero entropy and they correspond to solutions with no attractor behaviour: the scalars fields never reach a fixed point at finite distance in moduli space. Although the corresponding black hole solutions are singular, they have played an increasingly central role in recent literature because of their relation to  the issue of finiteness of N=8 supergravity \cite{Bianchi}.  Small Orbits were initially examined in \cite{FG}, then further discussed in \cite{orbits,reviewSF} while  in \cite{largesmall} they have been recently defined by the limit $I_4\to 0$ of large orbits. 

Beside the classification of large and small orbits, duality has also played a major role in finding  in some generality the {\sl fake} superpotential  $W$, which at first was only known within  particular  models \cite{Ceresole:2007wx,Andrianopoli:2007gt,Bellucci:2008sv,Cardoso5d,varinonBPS}. 
To begin with,  the $N=2$  central charge is a symplectic product of the symplectic  sections with the electric-magnetic charge vector $ Q=(p^\Lambda,q_\Lambda)$:  $Z=e^{K/2}(q_\Lambda X^\Lambda-p^\Lambda F_\Lambda)$.  Also the effective potential itself  (\ref{pot}) is a symplectic invariant quantity,  and the supersymmetric flow equations  (\ref{flows}) are also driven by the invariant  quantity $|Z|$.  Therefore, it is reasonable that also the {\sl fake} superpotential $W$ for the non-BPS branch be built out of  symplectic invariants, in agreement with \cite{Andrianopoli:2009je} where $W$ is interpreted as the Hamilton's principal function associated to the non-BPS flow equations.  A constructive strategy together with an explicit formula for  W in terms  of  duality invariants was given  in \cite{Ceresole:2009iy} for the $t^3$ model and more generally in \cite{universality} for the $st^2$ and $stu$ models, also in relation with the $N=8$ theory in the alternative approach of nilpotent orbits \cite{Bossard:2009we}.

In this note we elucidate the properties of the singular black hole solutions in four dimensional  $N=2$ supergravity coupled to $n$ vector multiplets, where the geometry of moduli space is given by special geometry\cite{Strominger:1990pd} and the duality is encoded in $Sp(2n+2)$ symplectic transformations.  Interestingly, although singular solutions have zero entropy and no attractor behaviour, one can still define W\cite{largesmall}, which has a runaway behaviour and gets stabilized only at the boundary of moduli space, by taking an $I_4\to 0$ limit of the $W$ for large orbits \cite{largesmall}. We will then  provide explicit forms for the fake superpotential W for all small orbits in $N=2$ symmetric theories.
Our goal here is to describe the amount of supersymmetry preserved in each small orbit and to find the relevant fake superpotential  W for each of them.   
After revisiting the large orbits, we intend to use this universal description in terms of invariants also towards the classification of   orbits of the $N=2$ charge vector for symmetric special geometries G/H, extending similar results obtained in \cite{Ferrara:1997ci} for maximally extended theories.

Duality invariant quantities are those that remain unchanged (transform as scalars) under the simultaneous action of the duality group on the charge vector $Q = (p^\Lambda, q_\Lambda)$ and on the scalar fields (expressed through the symplectic sections $(X^\Lambda,F_\Lambda)$, with $\Lambda=(0,i)=0,\ldots,n$). Here we recall that the complete set of  local H invariants in $N=2$  special geometry found in \cite{Cerchiai:2009pi} is given by
\begin{eqnarray}
	i_1 &=& Z{\overline {Z}}\label{i1}\\
 i_2 &=& g^{i{\bar\jmath}} Z_i{\overline {Z}}_{\bar\jmath}\,\qquad\qquad\qquad\qquad\quad (Z_i=D_iZ\ \, , \  {\overline {Z}}_{\bar\imath}={\overline {D}}_{\bar\imath}\ {\overline {Z}} )\, , 
\label{i2}\\
	i_3&=&\frac16\left[ Z N_3({\overline {Z}})+{\overline {Z}} {\overline N}_3 (Z_i)\right], \qquad i_4 = \frac{i}{6}\left[ Z N_3({\overline {Z}}) - {\overline {Z}} {\overline N}_3(Z)\right]\,, \label{i4}\\
	i_5 &=& g^{i{\bar\imath}}C_{ijk}C_{{\bar\imath}{\bar\jmath}\bar k}{\overline {Z}}^j{\overline {Z}}^k\, Z^{\bar\jmath} Z^{\bar k} \, , \label{i5}
\end{eqnarray}
where the cubic norms are given by
\begin{equation}
	N_3({\overline {Z}})=C_{ijk}{\overline {Z}}^i\ {\overline {Z}}^j\ {\overline {Z}}^k\,,\qquad \ \qquad {\overline N}_3(Z)=C_{{\bar\imath}{\bar\jmath}\bar k}Z^{\bar\imath}\ Z^{\bar\jmath}\ Z^{\bar k} .
\end{equation}
These five invariants  in the case of symmetric special geometries are not unrelated, and although each one of them depends on the scalar fields and the charges,  they satisfy a constraint which involves the quartic  G invariant $I_4$ which is field independent: 
\begin{equation}
	I_4=(i_1-i_2)^2+4 i_4-i_5\, , \qquad \partial_\varphi I_4=0\, , \quad \partial_{\bar \varphi} I_4=0\, . 
	\label{I4general}
\end{equation}
The objects $(i_1,\ldots ,i_5)$ behave as scalar functions of the charges and the scalar fields under duality transformations.

Our main interest is to compute the superpotential W, as its value at radial infinity gives the ADM mass of the given black hole. In extended supergravity, the BPS bound states that 
\begin{equation}
M_{ADM}=W(\varphi_\infty,Q)\geq |z_h|
\end{equation}
where $z_h$ is the highest skew eigenvalue of the central charge $Z_{AB}$, which is saturated for BPS solutions. In the N=8 case, an interesting bound holds for the non-BPS orbits\cite{Bianchi,largesmall}
\begin{equation}
|z_h|^2<W_{\text{nonBPS}}^2\leq4|z_h|^2\, ,
\end{equation}
since $W^2\leq V_{BH}\leq 4 |z_h|^2$.
In the $N=2$ theory, $Z_{AB}=\epsilon_{AB}Z$ ($A,B=1,2$) and the highest eigenvalue coincides with the N=2 central charge.

According to the procedure  used for the $N=8$ theory in \cite{D'Auria:1999fa}, where the four skew eigenvalues of the central charge were put in correspondence with the four eigenvalues of a quartic polynomial, and following the reasonings of \cite{universality},  one can establish a correspondence between the basis $(i_1,\ldots, i_5)$ and the set $(i_1,\lambda_1,\lambda_2,\lambda_3,I_4)$ where $\lambda_i$ are the roots of a universal cubic equation
\begin{equation}
\lambda^3-i_2 \lambda^2 +\frac{i_5}{4}\lambda-\frac{i_3^2+i_4^2}{4i_1}=0\,
\label{cubicequation}
\end{equation}
with real, positive roots  given by
\begin{eqnarray}
\lambda_1 &=&\frac{1}{3}\left(i_2+ 2\, {\rm Re}\, w\right), \label{sollambda1}\\
\lambda_2 &=&\frac{1}{3}\left(i_2-{\rm Re}\, w-\sqrt{3}\,{\rm Im}\, w\right),\\
\lambda_3 &=&\frac{1}{3}\left(i_2-{\rm Re}\, w+\sqrt{3}\,{\rm Im}\, w\right) . \label{sollambda3}
\end{eqnarray}
where
\begin{eqnarray}
v&=&2i_2^3+\frac{27(i_3^2+i_4^2)}{4i_1}-\frac{9i_2i_5}{4}\, ,\\
z&=&\frac{9i_2(i_3^2+i_4^2)i_5}{8i_1}+\frac{i_2^2i_5^2}{16}-\frac{i_2^3(i_3^2+i_4^2)}{i_1}-\frac{27(i_3^2+i_4^2)^2}{16i_1^2}-\frac{i_5^3}{16}\, ,\\
w&=&\left(\frac{v+3i\sqrt{3z}}{2}\right)^{1/3}\, .
\end{eqnarray}
 %where $v\geq 0$, and $z\geq 0$. 
Therefore $V_{BH}=i_1+i_2=i_1+\lambda_1+\lambda_2+\lambda_3$.
In a generic situation, the highest root of the cubic is $\lambda_1$ and for small black holes it will coincide with the fake superpotential $W_{\text{nonBPS}}$ in various non-BPS orbits to be discussed below.

%{\bf{Attractors and large orbits revisited}}

We start by giving a characterization of the attractors and the large orbits purely in terms of the invariants (\ref{i1})-(\ref{i5}), together with their W superpotential. For the BPS branch, we always have  $W^2=i_1$, while in the non BPS case it has to be determined case by case. We shall see that the only branch where W is not given by simple radicals is when $I_4<0$.

\noindent
The black hole  attractors in N=2 theories are defined as solutions of the equation \cite{FGK} 
\begin{equation}
\partial_i V_{BH}=2 {\overline Z} Z_i +i C_{ijk}{\overline Z}^j{\overline Z}^k=0\, .\label{AE}
\end{equation}
According to the classification of \cite{classN=2}, symmetric special geometries subdivide into various series. The first is given by the four Magic supergravities or  irreducible Jordan models (based on Jordan algebras over $\mathbb{O}$, $\mathbb{H}$, $\mathbb{C}$, $\mathbb{R}$ \cite{Jordan}).  The corresponding BPS attractor point, with $Z_i=0$ is given by
\begin{equation}
\text{BPS}:\qquad i_2=i_3=i_4=i_5=0\, ,\ \lambda_1=\lambda_2=\lambda_3=0\, , \ \ I_4=i_1^2\, , \quad W=\sqrt{i_1}\, .
\end{equation}
\noindent
They also admit a non BPS attractor point for $Z_i\neq 0$ and $Z = 0$, which  is given by
\begin{equation}
 \text{non BPS},\  Z=0\,:\quad i_1 = i_3=i_4 =i_5=0\, ; \quad\lambda_1=i_2,\ \lambda_2=\lambda_3=0;\quad I_4 = i_2^2\, , W=\sqrt{i_2}\, .
\end{equation}
Finally, a $Z_i\neq 0$, $Z \neq 0$ nonBPS attractor point occurs  at
\begin{equation}
	i_2 = 3 i_1, \quad i_3 =0, \quad i_4 = -2 i_1^2, \quad i_5 = 12 i_1^2\, ,
\end{equation}
and then at these points  
\begin{equation}
	 i_1=\lambda_1=\lambda_2=\lambda_3\, ,\qquad I_4 = - 16 i_1^2 <0\, , W=2\sqrt{i_1} \, .
\end{equation}
\noindent
The above attractor points are actually part of  three regular orbits, with $I_4\neq 0$, which can be characterized as:
 \begin{eqnarray}
I_4 &>&0\quad  : \begin{cases}\text{ BPS}\quad &i_1>\lambda_1,\lambda_2,\lambda_3\\
\text{non BPS}\quad & \lambda_1>i_1,\lambda_2,\lambda_3
\end{cases}\nonumber\\
&& \\
I_4 &<&0\quad 
\text{non BPS}\qquad \lambda_1\neq\lambda_2\neq\lambda_3\, .
\end{eqnarray}
\noindent
While the superpotentials for $I_4>0$ are respectively $W_{BPS}=\sqrt{i_1}$ and $W_{nonBPS}=\sqrt{\lambda_1}$, for $I_4<0$ it has a complicated expression involving radicals that has been found  by studying on the $st^2$ and the $stu$ models \cite{universality,Bossard:2009we}.

Another instance of symmetric special geometry is the infinite reducible series, 
\begin{equation}
N=2:\qquad\frac{G}{H}=\frac{SL(2,{\mathbb R})}{U(1)}\times\frac{SO(2,n)}{SO(2)\times SO(n)}\, \label{infseries}
\end{equation}  
where due to the factorization of the moduli space $Z_i=Z_s,Z_I$ and  the basic cubic equation has factorized eigenvalues $i_s,\lambda_1,\lambda_2$
where $i_s=Z_s{\overline Z_s}$. The BPS attractor point is
\begin{equation}
i_s=0, \qquad\lambda_1=\lambda_2=0,\qquad I_4=i_1^2
\end{equation}
while the non BPS attractor points for $I_4>0$ are
\begin{equation}
Z=0\, :\quad \qquad W=\sqrt{i_s}, \qquad i_1=\lambda_1=\lambda_2=0\qquad I_4=i_s^2\, ,
\end{equation}
\begin{equation}
Z=0, Z_IZ^I=0\, :\quad \qquad W=\sqrt{\lambda_1}, \qquad i_1= i_s=\lambda_2=0\qquad I_4=(Z_I{\overline Z}^I)^2\, .
\end{equation}

There exist four large orbits:
\begin{eqnarray*}
&&I_4>0\ \ \begin{cases} \text{BPS}\quad & i_1>i_s,\lambda_1,\lambda_2,\qquad W_{\text{BPS}}=\sqrt{i_1}\\ 
                                 \text{nonBPS}\quad &i_s>i_1,\lambda_1,\lambda_2,\qquad W_{\text{nonBPS}}=\sqrt{i_s}\\
                                 \text{nonBPS}\quad &\lambda_1>i_1,i_s,\lambda_2,\qquad W_{\text{nonBPS}}=\sqrt{\lambda_1}
                                 \end{cases}\\
&&I_4<0 \  \ \ \ \ \ \  \text{nonBPS}
\end{eqnarray*}
where again there is a complicated fake superpotential for $I_4<0$ which, however, on the curve $i_3=0$,  reduces to $W=\frac12(\sqrt{i_1}+\sqrt{\lambda_1}+\sqrt{\lambda_2}+\sqrt{\lambda_3})$ \cite{Andrianopoli:2007gt,universality,Bossard:2009we}.

{\section{ Small orbits and their W superpotential}}
We first consider small orbits for the 4 irreducible Magic supergravities  (based on Jordan algebras over $\mathbb{O}$, $\mathbb{H}$, $\mathbb{C}$, $\mathbb{R}$ \cite{Jordan}) and  we impose suitable differential constraints along the lines of \cite{Ferrara:1997ci,Cerchiai:2009pi}. The rank $r$ refers to the minimal number of charges that characterize the given orbit \cite{orbits}.

\vspace{10pt}
\noindent
{\sl i) Lightlike orbits}  

\noindent
They are simply defined by $I_4=0$ and have rank $r=3$. This condition allows to eliminate the invariant $i_5$ by posing $i_5=(i_1-i_2)+4i_4$. The basic cubic equations  (\ref{cubicequation}) in this case will have three distinct roots, assuming a specific (not particularly illuminating) form upon eliminating $i_5$. The BPS orbit will arise when $i_1>\lambda_1$,  while the non-BPS orbit will take place when $i_1<\lambda_1$. Therefore we will have
\begin{equation*}
I_4=0\ \ r=3:\qquad\begin{cases}\text{BPS} \quad i_1>\sqrt{\lambda_1};  &W_{\text{BPS}}=\sqrt{i_1} \\
                                             \text{nonBPS}\quad  i_1<\sqrt{\lambda_1};     &W_{\text{non BPS}}=\sqrt{\lambda_1} \end{cases}\, .
\end{equation*}

This is also in agreement with the analysis of \cite{largesmall} where this orbit is seen to arise as a limit $I_4\to 0$ of the $I_4>0$ non-BPS large orbit.

\vspace{10pt}
\noindent
{\sl ii) Critical Orbits}
 
\noindent
This second class of orbits is obtained by taking vanishing first derivatives of (\ref{I4general}) with respect to the central charges $Z$ and their covariant derivatives $Z_i$, using the definitions (\ref{i1}-\ref{i5}). It has rank $r=2$.  We find the system of equations

\begin{eqnarray}
I_4 &=&(i_1-i_2)^2+4i_4-i_5=0\, ,\nonumber\\
Z\frac{\partial I_4}{\partial Z}&=& 2(i_1-i_2)i_1+2(i_4+i \, i_3)=0\, , \\
Z_i \frac{\partial I_4}{\partial Z_i} &=&2(i_2-i_1)i_2+6(i_4-i\, i_3)-2i_5=0\, ,\nonumber
\end{eqnarray}
yielding the conditions
\begin{equation}
\partial I_4=0 \rightarrow 
\begin{cases}
  &i_3=0  \\
  &i_4=(i_2-i_1)i_1>0   \\
  &i_5=(i_2-i_1)(i_2+3i_1) \quad\quad (i_2>i_1)\, .
\end{cases}
\end{equation}
For the roots of the basic cubic equation, these  constraints on the first derivatives yield
\begin{equation}
z=0\, ,\ \ \ v=\frac14(3i_1-i_2)^3\, ,\ \ \ w=\left(\frac{v}2\right)^{1/3}\, .
\end{equation}
There are  two cases according to the possible values of the parameter $v$:

a) $i_1>\frac{i_2}{3}$, $v=v^*$, $w=w^*=\frac{3i_1-i_2}2$, so that $\lambda_1=i_1$ and $\lambda_2=\lambda_3=\frac{i_2-i_1}2$

b) $i_1<\frac{i_2}{3}$, $v\neq v^*$, $w=(\frac12+i\frac{\sqrt{3}}2)(\frac{i_2-3i_1}2)$, so that $\lambda_1=\frac{i_2-i_1}2$, $\lambda_2=i_1$ and $\lambda_3=\lambda_1$.

\noindent
The first is the BPS case, where the highest eigenvalue is $\lambda_1=i_1$, while the second case is the non BPS, where we find $\lambda_1=\frac{i_2-i_1}2$. Therefore we get
\begin{equation*}
\partial I_4=0\, , \  r=2:\qquad\begin{cases}\text{BPS}\quad i_2>i_1>\frac{i_2}3    & W_{\text{BPS}}=\sqrt{i_1} \\
 &i_1=\lambda_1;\lambda_2=\lambda_3=\frac{i_2-i_1}2\\
                          \text{nonBPS}\quad   i_1<\frac{i_2}3                 &W_{\text{non BPS}}=\sqrt{\frac{i_2-i_1}2} 
                          \end{cases}\, .
\end{equation*}

\vspace{10pt}
\noindent
{\sl iii) Doubly critical orbit}

\noindent
We must consider the projection of the second derivatives on the adjoint representation of G, $\partial^2_{Adj} I_4=0$, which, analogously to the N=8 and N=4 theories \cite{Ferrara:1997ci,Cerchiai:2009pi}, leads to two second order differential operators on $I_4$ that read: 
\begin{equation}
\left(C_{ijk}\frac{\partial^2}{\partial {Z_j}\partial {Z_k}}+2 i g_{i\bar\jmath}\frac{\partial^2}{\partial Z\partial {{\overline Z}_{\bar\jmath}}}\right)I_4=0\, ,\qquad
\left(R_{i\bar \jmath k\bar l}\frac{\partial^2}{\partial {Z_k}\partial {\overline{ Z_{\bar l}}}}+2g_{i\bar\jmath}\frac{\partial^2}{\partial{Z}\partial{\overline Z}}\right)I_4=0\, .
\end{equation}
\noindent
The first one results in an equation like the attractor equation (\ref{AE}), but with an opposite relative sign. Together, they yield the extra condition $i_2=3 i_1$ and thus $i_4=2i_1^2$ and $i_5=12 i_1^2$. Adding these constraints to the previous ones leads to only one BPS orbit, since one has $i_1=\lambda_1=\lambda_2=\lambda_3$.  We have
\begin{equation*}
\partial^2_{Adj}I_4=0\, ,\quad r=1:\quad\text{BPS}\quad i_1=\lambda_1=\lambda_2=\lambda_3;   \quad W_{\text{BPS}}=\sqrt{i_1}
\end{equation*}
In this case, the relations among $i_3,i_2,i_5$ are the same as in the $I_4<0$ attractor point, but with a flipped sign in  $i_4$.

To summarize,  the generic four Magic models of N=2 symmetric special geometries admit 5 small orbits, defined by three main classes of G-invariant constraints given in terms of the H-invariants $(i_1,i_2,i_3,i_4,i_5)$:
\begin{eqnarray}
1)  &r=3& \quad i_5=(i_1-i_2)^2+4i_4\\
2)   &r=2& \quad i_3=0\, ;\quad i_4=i_1(i_2-i_1)>0\, ;\quad  i_5=(i_2-i_1)(i_2+3i_1)\\
3)  &r=1& \quad i_2=3i_1\, ;i_3=0\, ;\quad i_4=2i_1^2\, ;\quad i_5=12 i_1^2\, .
\end{eqnarray}

%{\bf{The infinite cubic sequence}}
Among the allowed symmetric spaces for $N=2$ special geometry classified in \cite{classN=2} there is  the  infinite cubic sequence (\ref{infseries}) which describes the coupling of an arbitrary number $n$ of vector multiplets.  The factorization of the manifold into two elements requires a separate analysis. However, due to the similarity with its  N=4 ancestor
\begin{equation}
N=4:\qquad\frac{G}{H}=\frac{SL(2,{\mathbb R})}{U(1)}\times\frac{SO(6,n)}{SO(6)\times SO(n)}\,
\end{equation}
one can adapt the formalism of \cite{Andrianopoli:1997pn,Cerchiai:2009pi} to the present case by replacing $Z_{AB}$ by  the pair $(Z,i{\overline Z}_s)$ , where $s$ is the single modulus of $SL(2)/U(1)$. 
The unique quartic invariant combination which does not depend on the scalar fields $s,\varphi^I$, is given by
\begin{equation}
I_4 = S_1^2-S_2{\overline S_2}\, ,\label{s1s2}
\end{equation}
where now $S_1$ and $S_2$ must be given by  
\begin{eqnarray}
S_1 &=&|Z|^2+|Z_S|^2-Z_I{\overline Z}^I\, \qquad (Z_S=D_S\, Z\, ,Z_I=D_I Z)\, ,\\
S_2&=&2i\, Z{\overline Z}_s-Z_IZ^I\, . 
\end{eqnarray}
Then the quartic invariant (\ref{s1s2}) becomes
\begin{equation}
I_4 
%&=& S_1^2-S_2{\overline S_2}=\\&=&
=(Z{\overline Z}-Z_s{\overline Z_s}-Z_I{\overline Z^I})^2+
2i(Z{\overline Z_s}{\overline Z^I}{\overline Z_I}-{\overline Z}Z_s Z^IZ_I)
-4Z_s{\overline Z_s}Z^I{\overline Z_I}-Z_I Z^I{\overline Z_K}{\overline Z^K}
\end{equation}
which agrees with (\ref{I4general}) upon using, for this reducible manifold the invariants
\begin{eqnarray}
i_1 &=&Z{\overline Z}\qquad\qquad i_2 =Z_s{\overline Z_s}+Z_I{\overline Z^I}\\
i_4 &=&\frac{i}2 (Z{\overline Z_s}{\overline Z^I}{\overline Z_I}-{\overline Z} Z_sZ^IZ_I)\\
i_5 &=&4Z_s{\overline Z_s}Z^I{\overline Z^I}+Z_IZ^I{\overline Z_K}{\overline Z^K}
\end{eqnarray}
The cubic polynomial in this case factorizes as \cite{universality}
\begin{eqnarray}
&&(\lambda-i_s)(\lambda^2-a\lambda+b)=0\\
&&i_s=Z_s{\overline Z}_s\qquad a=Z_I{\overline Z}^I=i_I\qquad b=\frac14|Z_IZ^I|^2
\end{eqnarray}
which is consistent with the full cubic equation (\ref{cubicequation}) as
\begin{equation}
i_2=i_s+a=|Z_s|^2+Z_I{\overline Z}^I\, ,\qquad\frac{i_5}4=i_s i_I+b\, ,\qquad\frac{i_3^2+i_4^2}{4i_1}=i_s b\, .
\end{equation}

Similarly to the Magic supergravities, one can obtain the following classification of orbits (the nomenclature is taken from \cite{Cerchiai:2009pi,largesmall}) with $I_4=S_1^2-S_2 {\overline S_2}=0$  in terms of  $S_1$ and $S_2$ with 
\begin{equation}
S_1=i_1+i_s-i_I\, ,\qquad |S_2|= \left[i_5-4i_4+4i_s(i_1-i_I)\right]^{1/2}
\end{equation}

\begin{itemize}
\item{Lightlike orbit: $I_4=0$, $r=3$ :}
\begin{equation*}
\begin{array}{lllll}
%\hline
 %& & & &\\
C_1:&
                                \begin{cases} \text{BPS} & \text{if $i_1> i_s,\lambda_1,\lambda_2$}\\
                                                       \text{nonBPS} &\text{if $i_s> i_1,\lambda_1,\lambda_2$}
              \end{cases}  &\quad S_1>0 \quad& \text{Adj}_{SO(2,n)}\neq 0 &\  \text{Adj}_{SL(2)}\neq 0\ \\
%& & & & \\
%\hline
%& & & & \\
C_2:&\ \ \text{nonBPS}\quad \lambda_1>i_1,i_s,\lambda_2  &\quad S_1<0 \quad& \text{Adj}_{SO(2,n)}\neq 0 &\ \text{Adj}_{SL(2)}\neq 0\ \\
%& & & &  \\
%\hline
\end{array}
\end{equation*}

\item{Critical orbit: $\partial I_4=0$, $r=2$:}

\begin{equation*}
%\begin{array}{|c|c|c|c|c|}
\begin{array}{llll}
%\hline
% & & & &\\
A_1:&
\text{BPS}\quad i_1=i_s>\lambda_1=\lambda_2 &\quad S_1>0 \quad& \text{Adj}_{SO(2,n)}= 0\, \ ,\text{Adj}_{ SL(2)}\neq 0\ \\
%& & & & \\
%\hline

% & & & &\\
A_2:&
\text{non BPS}\quad \lambda_1=\lambda_2>i_1=i_s&\quad S_1<0 \quad& \text{Adj}_{SO(2,n)}= 0\, ,\  \text{Adj}_{SL(2)}\neq 0\ \\
%& & & & \\
%\hline
%& & & & \\
B: \ & \begin{cases} \text{BPS} & \text{if $i_1=\lambda_1> i_s=\lambda_2$}\\
                      \text{non BPS} &\text{if $i_s=\lambda_1> i_1=\lambda_2$}
\end{cases}                       &\quad S_1=S_2=0 \quad& {\text{Adj}_{SO(2,n)}\neq 0}\, ,\ \text{Adj}_{SL(2)}= 0\ \\
%& & & &  \\
%\hline
\end{array}
\end{equation*}

\item {Doubly critical $\partial^2_{{Adj}} I_4=0$, $r=1$:}

\begin{equation*}
\begin{array}{llll}
%\hline
% & & & &\\
A_3:&
\text{BPS}\quad i_1=i_s=\lambda_1=\lambda_2&\quad S_1=S_2=0 \quad& \text{Adj}_{SO(2,n)}=\text{Adj}_{SL(2)}= 0\ \\
%& & & & \\
%\hline

 \end{array}
\end{equation*}

\end{itemize} 

%\end{equation}
\noindent 
We see that reducibility of the manifold leads to a degeneration of orbits that turn out to be 8 rather than 5 of the Magic models case.  The splitting of the B and $C_1$ orbits compared to the $N=4$ case are related to the two possible situations $i_1>i_s$ and $i_i<i_s$ of the two eigenvalues of N=4 that here are not on the same footing.

As far as the superpotential W is concerned, we always have for the supersymmetric orbits in N=2 $
W_{\text{BPS}}=\sqrt{i_1}$, while for the non supersymmetric case we have
\begin{equation}
\begin{cases}
B:\quad&W_{\text{nonBPS}}=\sqrt{\lambda_1}=\sqrt{i_s};\\
A_2:\quad &W_{\text{nonBPS}}=\sqrt{\lambda_1}=\sqrt{\lambda_2};\\
C_1:\quad &W_{\text{nonBPS}}=\sqrt{i_s};\\
C_2:\quad      &W_{\text{nonBPS}}=\sqrt{\lambda_1}\, .
\end{cases}
\end{equation}
Note that the corresponding orbits $B$, $C_1$ in N=4 were 1/4BPS because in that case $i_s$ was the second eigenvalue of the central charge and there is a symmetry in the exchange $i_1\to i_s$.

%{\sl {The quadratic series}}
\vspace{15pt}
To complete the analysis of N=2 symmetric special geometries, one should still consider the quadratic series
\begin{equation}
\frac{G}{H}=\frac{SU(1,n)}{SU(n)\times U(1)}
\end{equation}
 having $C_{ijk}=0$ and then $i_3=i_4=i_5=0$.  For these values there is only one eigenvalue of the cubic, $\lambda=i_2$.
This is an interesting case that cannot be derived from five dimensions. The quartic invariant actually becomes quadratic,
\begin{equation}
\text {Quadratic series:}\qquad I_4=I_2=(i_1-i_2)^2\, ,
\end{equation}
and one is left with the simple analysis.  For the large orbits, one has \cite{orbits} :
\begin{equation}
\begin{cases}
\text{BPS}: i_1>i_2, \quad I_2>0,\qquad &W=\sqrt{i_1}\\
\text{non BPS}: i_1<i_2,\quad I_2<0\qquad & W=\sqrt{i_2}
\end{cases}
\end{equation}
with attractor points $i_2=0$ and $i_1=0$ respectively.

\noindent
There is only one small orbit, arising for $i_1=i_2$ everywhere; it has $W=\sqrt{i_1}$ and it is  BPS.

{\section{Summary}}
This  approach gives a clean intrinsic classification of both large and small orbits of $N=2$ black holes for special geometries based on symmetric spaces $G/H$ entirely in terms of the H invariants. This allows to make no reference to a particular symplectic frame or to special coordinates. Moreover, this formalism makes it transparent to  see where the various BPS conditions come from. As a further outcome, we have determined invariant expressions for the the fake superpotential $W_{\text{non BPS}}=\sqrt{\lambda_1}$ for each distinct small non supersymmetric orbit, which turns out to be always given by simple radicals of the highest root of the cubic. The only case where the fake superpotential is not given by a simple radical is the  large orbit $I_4<0$  discussed in \cite{universality}. The reason for the calculability of W on all the small orbits can be understood by looking at the $N=8$ theory, whose truncation to N=2 reproduces almost all the N=2 symmetric models. In N=8, all the small orbits are BPS, as they preserve 1/8, 1/4, 1/2 supersymmetry. Since the fake superpotential at infinity yields the ADM mass, the BPS bound implies
\begin{equation}
W_{\infty}>\{|z_1|,|z_2|,|z_3|,|z_4|\}
\end{equation}
where $\{ z_i\}$ are the four skew eigenvalues of the central charge matrix $Z_{AB}$. In the $N=2$ truncation, these four eigenvalue split into $\{Z,\lambda_1,\lambda_2,\lambda_3\}$, where $|Z|^2=i_1$. Then clearly for  small BPS black holes the BPS bound gives $M_{ADM}^2=|Z|>\lambda_1,\lambda_2,\lambda_3$, while for small  non BPS black holes $M_{ADM}^2=\lambda_1>|Z|,\lambda_2,\lambda_3$.  Conversely, for the large non BPS black holes $W_{\text{nonBPS}}$ must exceed the BPS bound and it must be greater than any of the eigenvalues $\{Z,\lambda_1,\lambda_2,\lambda_3\}$. Its expression must be computed by other means 
\cite{Ceresole:2009iy,Bossard:2009we,universality} and one can see by taking its value at the attractor point that  $W_{nonBPS}=2\sqrt{i_1}$  ($i_1=\lambda_1=\lambda_2=\lambda_3$) which indeed is double the value of the BPS case. 

As a last comment, the relation between invariants $I_4$ and $I_3$ suggests that one can use this method to invariantly describe the stratification of orbits between five and four dimensions, a task that is left for future work.
\vspace{15pt}
\section*{Acknowledgments}
This work is supported in part by the ERC Advanced Grant no. 226455, \textit{``Supersymmetry, Quantum Gravity and Gauge Fields''} (\textit{SUPERFIELDS}) and in part by MIUR-PRIN contract 20075ATT78 and DOE Grant DE-FG03-91ER40662. The work of A. M. has been supported by an INFN visiting Theoretical Fellowship at SITP, Stanford University, CA, USA.

\end{document}